\documentclass{appolb-blank}  

\usepackage{graphicx}
\usepackage{latexsym,amssymb,amsmath,amsfonts,epsfig,mathtools}  

\usepackage[T1]{fontenc} 

\newcommand{\beq}{\begin{equation}}   
\newcommand{\eeq}{\end{equation}}
\newcommand{\beqa}{\begin{eqnarray}}
\newcommand{\eeqa}{\end{eqnarray}}
\newcommand{\beqNO}{\begin{equation*}}
\newcommand{\eeqNO}{\end{equation*}}
\newcommand{\beqaNO}{\begin{eqnarray*}}
\newcommand{\eeqaNO}{\end{eqnarray*}}
\newcommand{\bsubeqs}{\begin{subequations}}
\newcommand{\esubeqs}{\end{subequations}}
\newcommand\ddfracNEW[2]{\displaystyle{\frac{#1}{#2}}}

\begin{document}

\eqsec

\title{M-theory and the birth of the Universe%
\thanks{Presented at the XXVIIth Cracow Epiphany Conference
on the Future of Particle Physics, Cracow, Poland,
January 7--10, 2021.}%
}
\author{F.R. Klinkhamer
\address{
\begin{center}
Institute for Theoretical Physics,
Karlsruhe Institute of Technology (KIT),\\
76128 Karlsruhe, Germany\\
\texttt{frans.klinkhamer@kit.edu}
\end{center}
}
}
\maketitle
\begin{abstract}

\begin{center} 
\vspace*{-1\baselineskip}
\textit{In memory of Martinus J. G. Veltman (1931-2021)}\\
\end{center}
\vspace*{1.5\baselineskip}\noindent
In this review article, we first discuss a possible
regularization of the big bang curvature singularity of the
standard Friedmann cosmology, where the curvature singularity is
replaced by a spacetime defect.
We then consider the hypothesis that a new physics phase
gave rise to
this particular spacetime defect.
Specifically, we set out on
an explorative calculation using the IIB matrix model,
which has been proposed as a particular formulation
of nonperturbative superstring theory (M-theory).
\end{abstract}

\vspace{-123mm}
\noindent Acta Phys. Pol. B 52, 1007 (2021) \hfill arXiv:2102.11202
\vspace{+123mm}



\section{Introduction}
\label{sec:Intro}

As regards the topic of this conference, we submit
that the future of particle physics may very well
involve the future of gravitation and cosmology.
The title of the present contribution is in that
spirit, as ``M-theory'' concerns a possible future theory
merging elementary particle physics and
gravitational physics, while ``the birth of the Universe''
impacts on all topics of cosmology.

This review article presents one particular approach to the
question of how the Universe got started. First, we discuss a
possible regularization of the big bang curvature singularity,
while staying within the realm of general relativity but allowing
for degenerate spacetime metrics
(having, for example, a vanishing determinant of the metric
at certain spacetime points).
This regularization replaces, in fact,
the big bang curvature singularity
by a three-dimensional spacetime defect with
a locally vanishing determinant of the metric.

Next, we investigate the hypothesis that a new physics phase
produced this spacetime defect. In order to allow for
explicit calculations, we turn to the IIB matrix model,
which has been suggested as a formulation
of nonperturbative type-IIB superstring theory (M-theory).
The crucial question, now, is how
a classical spacetime might emerge from the IIB matrix model.
The answer appears to be that such an emerging classical spacetime
may reside in the large-$N$ master field
of the IIB matrix model. This master field
can, in principle, give the regularized-big-bang
metric of general relativity.
A word of caution is, however, called for:
as the formulation and validity of
M-theory have not yet been established,
the present review paper provides no definitive answers
but only a few suggestive results.

The outline is as follows.
In Sec.~\ref{sec:Standard-Friedmann-cosmology}, we recall
the main points of the standard Friedmann cosmology
and its big bang curvature singularity,
primarily to establish our notation.
In Sec.~\ref{sec:Regularized-big-bang}, we present a
particular regularization of the Friedmann big bang singularity
and highlight a few subtle points
(e.g., differential structure and singularity theorems).
In Sec.~\ref{sec:New-phase-from-M-theory}, we turn
to the main topic, namely the hypothetical existence of a
new physics phase giving rise to a classical spacetime
and possibly to a regularized (tamed) big bang.
Here, we rely on a particular matrix model realization
of M-theory, specifically the IIB matrix model.
Our focus is on the basic ideas and
technical details are relegated to four appendices.
In Sec.~\ref{sec:Conclusion}, we give a brief summary
and point out what the main outstanding task appears to be.

\section{Standard Friedmann cosmology}
\label{sec:Standard-Friedmann-cosmology}
\subsection{Robertson--Walker metric and Friedmann equations}
\label{subsec:Robertson-Walker-metric}

The Einstein gravitational field equation of general relativity
reads~\cite{Einstein1916}%
\begin{equation}
\label{eq:Einstein-eq}
R_{\mu\nu} - \frac12\, g_{\mu\nu}\,R =
- 8\pi G\;T_{\mu\nu}^\text{\,(M)} \,,
\end{equation}
with $R_{\mu\nu}$ the Ricci curvature tensor, $R$ the Ricci curvature
scalar, $T_{\mu\nu}^\text{\,(M)}$ the energy-momentum
tensor of the matter (described by the standard model of
elementary particles and possible extensions),
and $G$ Newton's gravitational coupling constant.
In this section and the next, we use relativistic units with $c=1$ and
the spacetime indices $\mu$, $\nu\,$ run over $\{0,\, 1,\, 2,\, 3 \}$.

For the record, we give
the energy-momentum tensor of a perfect fluid~\cite{Einstein1916}: 
\beq\label{eq:Tmunu-perfect-fluid}
T_{\mu\nu}^\text{\,(M,\;perfect\;fluid)}=
\left(P_{M}+\rho_{M}\right)\,U_{\mu}\,U_{\nu}+P_{M}\;g_{\mu\nu}\,,
\eeq
with a normalized four-velocity $U^{\mu}$ of the
comoving fluid element and scalars $\rho_{M}$ and $P_{M}$
(corresponding to the energy density and the pressure measured
in a localized inertial frame comoving with the fluid).

We will now review a special solution of the
Einstein equation with a relativistic perfect fluid,
namely the Friedmann--Lema\^{i}tre--Robertson--Walker
expanding universe~\cite{Friedmann1922-1924,Lemaitre1927,%
Robertson1935,Walker1937}.
For a homogeneous and isotropic cosmological model,
the spatially flat Robertson--Walker (RW)
metric is~\cite{Robertson1935,Walker1937}
\beq\label{eq:RW-ds2}
ds^{2}\,\Big|^\text{(RW)}
\equiv
g_{\mu\nu}(x)\, dx^{\mu}\,dx^{\nu} \,\Big|^\text{(RW)}
=
- d t^{2}+ a^{2}( t )\;\delta_{m n}\,dx^{m}\,dx^n\,,
\eeq
with $x^{0}=c\, t$ and $c=1$.
The spatial indices $m$, $n$  run over $\{1,\, 2,\, 3 \}$.

Consider a homogeneous perfect fluid
with energy density $\rho_{M}(t)$ and pressure $P_{M}(t)$.
Then, the Einstein equation \eqref{eq:Einstein-eq}
with the RW metric \textit{Ansatz} \eqref{eq:RW-ds2}
and the energy-momentum tensor of a homogeneous perfect
fluid \eqref{eq:Tmunu-perfect-fluid}
gives the following spatially flat Friedmann equations:%
\bsubeqs\label{eq:Feqs}
\beqa
\label{eq:1stFeq}
\hspace*{-0mm}&&
\left( \frac{\dot{a}}{a}\right)^{2}
= \frac{8\pi G}{3}\,\rho_{M}\,,
\\[2mm]
\label{eq:2ndFeq}
\hspace*{-0mm}&&
\frac{\ddot{a}}{a}+
\frac{1}{2}\,\left( \frac{\dot{a}}{a}\right)^{2}
=
-4\pi G\,P_{M}\,,
\\[2mm]
\label{eq:rhoMprimeeq}
\hspace*{-0mm}&&
\dot{\rho}_{M}+ 3\;\frac{\dot{a}}{a}\;
\Big(\rho_{M}+P_{M}\Big) =0\,,
\qquad
P_{M} = P_{M} \big(\rho_{M}\big)\,,
\eeqa
\esubeqs
where the overdot stands for differentiation with respect to $t$.
The first equation in \eqref{eq:rhoMprimeeq}
corresponds to energy conservation
and the second stands for the equation-of-state (EOS)
relation between pressure and energy density of the perfect fluid.
The matter is, moreover, assumed to satisfy
the standard energy conditions. The null energy condition
of the perfect fluid \eqref{eq:Tmunu-perfect-fluid}, for example,
corresponds to the inequality $\rho_{M}+P_{M} \geq 0$.

\subsection{Big bang singularity}   
\label{subsec:Big-bang-singularity}

The Friedmann equations \eqref{eq:Feqs}
for relativistic matter with constant EOS parameter
$w_{M} \equiv P_{M}/\rho_{M} =1/3$ 
give the well-known Friedmann--Lema\^{i}tre--Robertson--Walker (FLRW)
solution~\cite{Friedmann1922-1924,Lemaitre1927,Robertson1935,Walker1937}:%
\bsubeqs\label{eq:Friedmann-solution}
\beqa
\label{eq:Friedmann-asol}
a(t)\,\Big|^\text{(FLRW)}_{(w_{M}=1/3)} &=& \sqrt{t/t_{0}}\,,
\phantom{\rho_{M0}/a^{4}(t) =\rho_{M0}\; t_{0}^{2}/t^{2}}
\hspace*{-4mm}
\text{for}\;\;\;t>0\,,
\\[2mm]
\label{eq:Friedmann-rhoMsol}
\rho_{M}(t)\,\Big|^\text{(FLRW)}_{(w_{M}=1/3)} &=&
\rho_{M0}/a^{4}(t) =\rho_{M0}\; t_{0}^{2}/t^{2}\,,
\phantom{\sqrt{t/t_{0}}}
\hspace*{-4mm}
\text{for}\;\;\;t>0\,,
\eeqa
\esubeqs
where the cosmic scale factor has normalization
$a(t_{0})=1$ at $t_{0}>0$ and the constant $\rho_{M0}$ is positive
[from \eqref{eq:1stFeq}, we have that
$G \rho_{M0}$ is proportional to $1/t_0^{2}\,$].
This FLRW solution displays the big bang singularity for $t\to 0^{+}$:%
\beq
\lim_{t\to 0^{+}} a(t) =0\,,
\eeq
with diverging curvature
(e.g., a diverging Kretschmann curvature scalar
$K \equiv R^{\mu\nu\rho\sigma}\,R_{\mu\nu\rho\sigma}\propto 1/t^{4}$)
and diverging energy density \eqref{eq:Friedmann-rhoMsol}.

At $t=0$, however, the used theory  
(i.e., general relativity and the standard model of elementary
particles) is no longer valid
and we can ask what happens \emph{really}?  
Or, more precisely: \emph{how to describe the birth of the Universe?}

\section{Regularized big bang}
\label{sec:Regularized-big-bang}

\subsection{New metric and modified Friedmann equations}
\label{subsec:New-metric-and-modified-Friedmann-equations}

First, we set out to control the
divergences of the standard Friedmann cosmology or,
in other words, to ``regularize'' the big bang singularity.
We do this by using a
new \textit{Ansatz}~\cite{Robertson1935,Walker1937,Klinkhamer2019-rbb}:
\bsubeqs\label{eq:reg-bb}
\beqa\label{eq:reg-bb-ds2}
\hspace*{0.0mm}
ds^{2}\,\Big|^\text{(RWK)}
&\equiv&
g_{\mu\nu}(x)\, dx^{\mu}\,dx^{\nu} \,
\Big|^\text{(RWK)}
=  
- \ddfracNEW{t^{2}}{t^{2}+b^{2}}\,d t^{2}
+ a^{2}( t )
\;\delta_{m n}\,dx^{m}\,dx^{n}\,,
\nonumber\\  \hspace*{0.0mm}&&   
\\[2mm]
\hspace*{0.0mm}
b &>& 0\,,
\quad
a^{2}( t ) > 0\,,
\\[2mm]
\label{eq:ranges-cosmic-coordinates}
\hspace*{0.0mm}
 t    &\in& (-\infty,\,\infty)\,,\quad
 x^{m} \in (-\infty,\,\infty)\,,
\eeqa
\esubeqs
with $x^{0}=c\, t$ and $c=1$. The nonzero length scale $b$ enters
the metric component $g_{00}(t)$ and will be seen to act as a regulator.
Setting $b=0$ in the metric \eqref{eq:reg-bb-ds2} formally
reproduces the RW metric \eqref{eq:RW-ds2} with $g_{00}=-1$,
but this does not really hold for the metric \eqref{eq:reg-bb}
at $t=0$, which has $g_{00}(0)=0$. In short,
the limits $t \to 0$ and $b \to 0$ do not commute for $g_{00}(t)$
from \eqref{eq:reg-bb-ds2}.

The metric $g_{\mu\nu}(x)$ from \eqref{eq:reg-bb}
is \emph{degenerate}, with a vanishing determinant at $t = 0$.
Physically, the $t = 0$ slice corresponds
to a three-dimensional \emph{spacetime defect}
(the terminology has been chosen so as to emphasize
the analogy with a defect in an atomic crystal).
See Refs.~\cite{Klinkhamer2020-more,KlinkhamerWang2019-cosm,%
KlinkhamerWang2020-pert,Wang2020-PhD-thesis} for further discussion
of  the new cosmological metric  \eqref{eq:reg-bb}
and Refs.~\cite{Klinkhamer2014-prd,KlinkhamerSorba2014,%
Guenther2017,Klinkhamer2019-JPCS}
for some background on this type of spacetime defect.

At this moment, we have two general observations.
First, if we replace the symbol $t$ from \eqref{eq:RW-ds2}
by $\tau \in \mathbb{R}$ and perform surgery on $\tau$
(introducing the length scale $b$ by
removing the open interval between the points $\tau=-b$ and $\tau=+b$),
then there is a coordinate transformation~\cite{Klinkhamer2019-rbb}
between the resulting
\newline\noindent   
$\tau$ and $t \in \mathbb{R}$, which
transforms the metric from \eqref{eq:RW-ds2}
into the one from \eqref{eq:reg-bb-ds2}.
But this coordinate transformation is \emph{not} a diffeomorphism
(an invertible $C^{\infty}$ function),
as the two points $\tau=\pm b$
are mapped into the single point $t=0$.
This implies that the differential structure corresponding
to \eqref{eq:RW-ds2} is different from the one corresponding
to \eqref{eq:reg-bb-ds2}; see also Ref.~\cite{KlinkhamerSorba2014} and
the second remark at the end of this subsection for related comments.

Second, the metric \eqref{eq:reg-bb-ds2} at $t\sim b$
is a \emph{large} perturbation away from the
RW metric \eqref{eq:RW-ds2} for equal values of $a(t)$,
so that $t \gg b$ would be required
in \mbox{Hawking's} argument for the occurrence of a singularity
(see the sentence ``Therefore any perturbation $\ldots$''
from the top-left column on p. 690 of Ref.~\cite{Hawking1965}).
Anyway, more general cosmological singularity
theorems~\cite{Hawking1967,HawkingPenrose1970,Wald1984}
hold true and we shall comment on their interpretation
in the last paragraph of Sec.~\ref{subsec:Bounce-or-new-physics-phase?}.

The standard Einstein equation \eqref{eq:Einstein-eq}
with the new metric \textit{Ansatz} \eqref{eq:reg-bb}
and a homogeneous perfect fluid \eqref{eq:Tmunu-perfect-fluid}
gives \emph{modified} spatially flat Friedmann equations:%
\vspace*{-2.00mm}  
\bsubeqs\label{eq:mod-Feqs}
\beqa
\label{eq:mod-1stFeq}
\hspace*{-0mm}&&
\left[1+ \frac{b^{2}}{t^{2}}\,\right]\,
\left( \frac{\dot{a}}{a}\right)^{2}
= \frac{8\pi G}{3}\,\rho_{M}\,,
\\[2.00mm]  
\label{eq:mod-2ndFeq}
\hspace*{-0mm}&&
\left[1+\frac{b^{2}}{t^{2}}\,\right]\,
\left(\frac{\ddot{a}}{a}+
\frac{1}{2}\,\left( \frac{\dot{a}}{a}\right)^{2}
\right)
-\frac{b^{2}}{t^{3}}\,\frac{\dot{a}}{a}
=
-4\pi G\,P_{M}\,,
\\[2.00mm]  
\label{eq:mod-Feq-rhoMprimeeq}
\hspace*{-0mm}&&
\dot{\rho}_{M}+ 3\;\frac{\dot{a}}{a}\;
\Big(\rho_{M}+P_{M}\Big) =0\,,
\qquad
P_{M} = P_{M} \big(\rho_{M}\big)\,,
\eeqa
\esubeqs
\vspace*{-4.00mm}\newline  
where the overdot stands again for differentiation with respect to $t$.
Two remarks are in order.
First, the inverse metric from \eqref{eq:reg-bb-ds2}
has a component $g^{00}=(t^{2}+b^{2})/t^{2}$
that diverges at $t=0$  and we must be careful to obtain
the reduced field equations at $t=0$ from the limit $t \to 0$
(see Sec.~3.3.1 of Ref.~\cite{Guenther2017} for further
details and Ref.~\cite{Horowitz1991} for a discussion of the
practical advantages of using a first-order formalism).
Second, the new $b^{2}$ terms in
the modified Friedmann equations \eqref{eq:mod-1stFeq}
and \eqref{eq:mod-2ndFeq} are a manifestation of the
different differential structure of  \eqref{eq:reg-bb-ds2}
compared to the differential structure of \eqref{eq:RW-ds2}
which gives the standard Friedmann equations \eqref{eq:1stFeq}
and \eqref{eq:2ndFeq}.

\subsection{Regular solution}
\label{subsec:Regular-solution}

Having obtained modified Friedmann equations, it is clear that we
expect to get modified solutions. In fact,
for constant EOS parameter $w_{M} \equiv P_{M}/\rho_{M} =1/3$,
the even solution of \eqref{eq:mod-Feqs} 
reads~\cite{Friedmann1922-1924,Lemaitre1927,%
Robertson1935,Walker1937,Klinkhamer2019-rbb}
\bsubeqs\label{eq:regularized-Friedmann-asol-rhoMsol}
\beqa\label{eq:regularized-Friedmann-asol}
a(t)\,\Big|_{(w_{M}=1/3)}^\text{(FLRWK)}
&=&
\sqrt[4]{\big(t^{2}+b^{2}\big)\big/
\big(t_{0}^{2}+b^{2}\big)}\,,
\\[4mm]
\label{eq:regularized-Friedmann-rhoMsol}
\rho_{M}(t)\,\Big|^\text{(FLRWK)}_{(w_{M}=1/3)} &=&
\rho_{M0}/a^{4}(t) =\rho_{M0}\,
\big(t_{0}^{2}+b^{2}\big)\big/\big(t^{2}+b^{2}\big) \,,
\eeqa
\esubeqs
where the cosmic scale factor has normalization
$a(t_{0})=1$ at $t_{0}>0$ and the constant $\rho_{M0}$ is positive
[from \eqref{eq:mod-1stFeq}, we have
$G \rho_{M0} \propto 1/(b^{2}+t_0^{2})\,$].

The new solution \eqref{eq:regularized-Friedmann-asol-rhoMsol}
is perfectly smooth at $t=0$ as long as $b\ne 0$,
and the same holds for the corresponding
Kretschmann curvature scalar $K(t) \propto 1/(b^{2}+t^{2})^{2}$.
Figure~\ref{fig:a-bounce-a-singular}
compares this regularized FLRWK solution (full curve)
with the singular FLRW solution (dashed curve).

Observe that the function $a(t)$ from \eqref{eq:regularized-Friedmann-asol}
is convex over a \emph{finite} interval around $t=0$, whereas
the function $\widetilde{a}(\tau)$ from \eqref{eq:Friedmann-asol},
with $t$ replaced by $\tau$ and $t_{0}$ by $\tau_{0}$, is concave
for $\tau \geq b >0$.
This different behavior of $a(t)$ just above $t=0$ (convex)
and $\widetilde{a}(\tau)$ just above $\tau=b$ (concave)
results from the different differential structures mentioned at the end
of Sec.~\ref{subsec:New-metric-and-modified-Friedmann-equations}.

\subsection{Bounce or new physics phase?}
\label{subsec:Bounce-or-new-physics-phase?}

\begin{figure}[t]
\begin{center}
\hspace*{0mm}   
\includegraphics[width=0.55\textwidth]{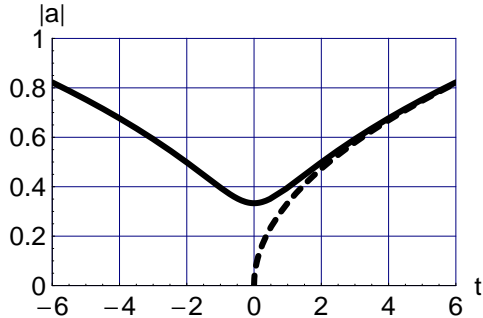}
\end{center}
\vspace*{-0mm}
\caption{Cosmic scale factor $a(t)$ of the spatially flat
FLRWK universe with $w=1/3$ matter (full curve),
as given by \eqref{eq:regularized-Friedmann-asol}
for $b=1$ and $t_{0}=4\,\sqrt{5}$.   
Also shown is the cosmic scale factor of the standard spatially flat
FLRW universe with $w=1/3$ matter (dashed curve),
as given by \eqref{eq:Friedmann-asol} with $t_{0}=4\,\sqrt{5}$.
\vspace*{-0mm}}
\label{fig:a-bounce-a-singular}
\end{figure}

With the regular solution \eqref{eq:regularized-Friedmann-asol-rhoMsol}
in hand, there are now two scenarios:
\begin{enumerate}
\vspace*{-0mm}
\item
A nonsingular bouncing cosmology, where the cosmic
time coordinate $t$ runs from $-\infty$ to $+\infty$.
This scenario may hold for $b \gg l_\text{Planck}$,
so that classical Einstein gravity can be expected to be applicable.
A possibly more realistic solution than the one
from \eqref{eq:regularized-Friedmann-asol-rhoMsol}
has EOS parameter $w_{M}=1$ in the prebounce epoch and $w_{M}=1/3$
in the postbounce epoch~\cite{KlinkhamerWang2019-cosm}.
Potential experimental signatures may rely on gravitational waves
generated in the
prebounce epoch, which keep on propagating into the postbounce
epoch~\cite{KlinkhamerWang2020-pert}.
\vspace*{-0mm}
\item
A new physics phase at $t=0$, which produces
two apparently similar universes.
This scenario may hold for $b \sim l_\text{Planck}$.
A special scenario has the new physics phase at $t=0$
pair-producing a ``universe'' for $t>0$ and an ``antiuniverse'' for $t<0$.
The relative role of particles and antiparticles in the $t > 0$
branch is then reversed compared to the one in the separate $t<0$ branch.
In order to obtain this particle-antiparticle behavior, it appears
necessary to use the (discontinuous) \emph{odd} solution
for $a(t)$, as given by the right-hand side of
\eqref{eq:regularized-Friedmann-asol} for $t > 0$
and the same with an overall minus sign for $t<0$.
(The full curve of Fig.~\ref{fig:a-bounce-a-singular}
also applies to this odd solution, as the figure plots
the absolute value of $a$.)
See Ref.~\cite{BoyleFinnTurok2018} for further discussion.
\vspace*{-1mm}
\end{enumerate}

\vspace*{0mm}

For both scenarios, the $t=0$ slice corresponds to
a three-dimensional \emph{spacetime defect}, which manifests itself
as a finite discontinuity at $t=0$ in the trace $K_\text{extr}(t)$ of
the extrinsic curvature on constant-$t$ hypersurfaces.
Also, there is a finite discontinuity at $t=0$ in the
expansion $\theta(t)$ for a particular congruence of timelike geodesics
(for the standard FLRW solution, $\theta(t)$ diverges as $t\to 0^{+}$;
see Sec.~4.2.3 of Ref.~\cite{Wang2020-PhD-thesis}
for the explicit expressions of $\theta$).
The discontinuous behavior of $K_\text{extr}(t)$ and $\theta(t)$,
in the spacetime with metric \eqref{eq:reg-bb},  
shows that $K_\text{extr}(t)$ and $\theta(t)$ are \emph{ill-defined} at $t=0$.
A related observation is that the $t=0$ hypersurface of the spacetime
with metric \eqref{eq:reg-bb} 
is not a Cauchy surface (cf. Sec.~10.2 of Ref.~\cite{Wald1984}).
%
%

It is unclear, for the first scenario, what physical mechanism
determines the relatively large value of $b$
and the \textit{raison d'\^{e}tre} of the spacetime defect
in the first place.
For the second scenario, the hope is that
the new physics sets the value of $b$ and also explains the origin
of the spacetime defect.
The focus of this review article is on the second scenario. 

In any case, there is no doubt as to the validity of
the Hawking and Hawking--Penrose cosmological
singularity theorems~\cite{Hawking1967,HawkingPenrose1970}
and the ``singularity'' of these theorems
may very well correspond to a three-dimensional spacetime defect
with a locally degenerate metric, as long as the
defect is explained by new physics (if, for example,
the defect is produced as a remnant of a new physics phase).
We refer the reader to the paragraph starting with
``This brings us to the third question: the nature of the singularity''
in Sec.~1 of Hawking's paper~\cite{Hawking1967}.
Our suggested new physics would, in principle, be physically observable,
which would  address Hawking's reservations
about including the degeneracy points in the definition of spacetime
(see, in particular, the sentence with the word ``undesirable''  in the
paragraph mentioned).

\section{New phase from M-theory}
\label{sec:New-phase-from-M-theory}

\subsection{Preliminary remarks}
\label{subsec:Preliminary-remarks}

We will now explore the idea that a new physics
phase gave rise to classical spacetime and matter,
as described by general relativity and the standard model
of elementary particles.
The fabric of classical spacetime would emerge from this
new physics phase and classical spacetime
might resemble an atomic crystal.
But, for an atomic crystal, we know that, if the formation of the crystal
is rapid enough, there may occur crystal defects.
By analogy, it could then be that the hypothetical new physics
phase produced the regularized-big-bang spacetime \eqref{eq:reg-bb},
with a three-dimensional spacetime defect.

\begin{figure}[t]
\vspace*{0mm}
\begin{center}
\hspace*{0mm} 
\includegraphics[width=0.65\textwidth]{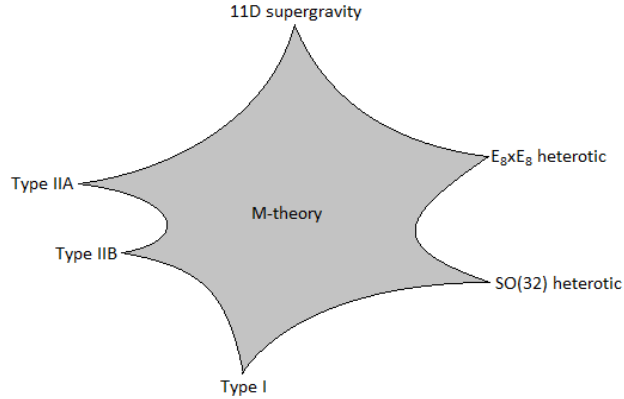}
\end{center}\vspace*{-0mm}
%
%
%
\caption{Sketch of the relationship between M-theory,
the five ten-dimensional \mbox{superstring} theories,
and eleven-dimensional supergravity theory.
\hfill License notice:\newline
\mbox{Polytope24 (https://commons.wikimedia.org/wiki/File:
\protect{\detokenize{Limits_of_M-theory.png}}),}\newline
``Limits of M-theory,'' https://creativecommons.org/licenses/by-sa/3.0/legalcode.
}
\label{fig:M-theory}
\end{figure}

For the moment, we do not know for sure how to describe
such a new physics phase. One candidate theory is M-theory.
Recall that M-theory is a hypothetical theory that
unifies all five consistent versions of superstring theory
(cf. Refs.~\cite{Witten1995,HoravaWitten1996,Duff1996});
see the ``nerve-cell'' sketch of Fig.~\ref{fig:M-theory}.

For an \emph{explicit} description of the
new physics phase replacing the big bang,
we will use the IIB matrix model
of Kawai and collaborators~\cite{IKKT-1997,Aoki-etal-review-1999},
which has been proposed as a nonperturbative formulation of
type-IIB superstring theory and, thereby, of M-theory
(the assumption is that all theories of Fig.~\ref{fig:M-theory}
belong to the same universality class).

\subsection{IIB matrix model}
\label{subsec:IIB-matrix-model}

The IIB matrix model, also known as IKKT
model~\cite{IKKT-1997,Aoki-etal-review-1999}, has a finite number
of $N \times N$ traceless Hermitian matrices:
ten bosonic matrices $A^{\mu}$ and
essentially eight fermionic (Majorana--Weyl)
matrices $\Psi_{\alpha}$.

The partition function $Z$ of the Lorentzian IIB matrix model
is defined by the following ``path''
integral~\cite{IKKT-1997,Aoki-etal-review-1999,KimNishimuraTsuchiya2012,%
NishimuraTsuchiya2019}:%
\bsubeqs\label{eq:IIB-matrix-model-ZwithSgeneral-S-eta}
\beqa
\label{eq:IIB-matrix-model-ZwithSgeneral}
\hspace*{-12mm}&&
Z =\int dA\,d\Psi\;\exp\left[i\,S/ \ell^{4}\,\right]
= \int dA\;\exp\left[i\,S_\text{eff}/ \ell^{4}\,\right]\,,
\\[2mm]
\label{eq:IIB-matrix-model-action}
\hspace*{-12mm}&&
S =  -\text{Tr}\,
\Bigg(\!
\frac{1}{4}\,\big[ A^{\mu} ,\,A^{\nu}    \big]\,
             \big[ A^{\rho},\,A^{\sigma} \big]\,
             \widetilde{\eta}_{\mu\rho}\,\widetilde{\eta}_{\nu\sigma}
+\frac{1}{2}\, \overline{\Psi}_{\beta}\,
\widetilde{\Gamma}^{\mu}_{\beta\alpha}\,\widetilde{\eta}_{\mu\nu}
 \,\big[ A^{\nu},\,\Psi_{\alpha} \big]
\!\Bigg),
\\[2mm]
\label{eq:IIB-matrix-model-etamunu}
\hspace*{-12mm}&&
\widetilde{\eta}_{\mu\nu} =
\Big[ \text{diag}
\big(  -1,\,  1,\, \ldots \,,\,1 \big)
\Big]_{\mu\nu}\,,
\;\;\;\text{for}\;\;\;
\mu, \nu \in \{0,\, 1,\, \ldots\, ,\, 9 \}\,,
\eeqa
\esubeqs
where the action \eqref{eq:IIB-matrix-model-action} contains only
Yang--Mills-type commutators, defined
by $[X,\,Y]\equiv$ $X \cdot Y - Y \cdot X$  
for square ma\-tri\-ces $X$ and $Y$.
The fermionic matrices have been integrated
out in the last step of \eqref{eq:IIB-matrix-model-ZwithSgeneral}
and the effective action $S_\text{eff}$ includes a complicated 
high-order term in addition to the bosonic quartic term. 
Expectation values of further observables will be discussed later.

We have two technical remarks. First, we speak of the
``Lorentzian'' IIB matrix model, because the
coupling constants $\widetilde{\eta}_{\mu\nu}$
in \eqref{eq:IIB-matrix-model-etamunu} resemble the Lorentzian metric
of ten-dimensional Minkowski spacetime.
Second, a model length scale ``$\ell$'' has been introduced
in \eqref{eq:IIB-matrix-model-ZwithSgeneral}, so that
$A^{\mu}$ from \eqref{eq:IIB-matrix-model-action}
has the dimension of length and $\Psi_{\alpha}$
the dimension of $(\text{length})^{3/2}$.

Now, the matrices $A^{\mu}$ and
$\Psi_{\alpha}$ in \eqref{eq:IIB-matrix-model-ZwithSgeneral}
are merely \emph{integration variables}.
Moreover, there is
\emph{no obvious small dimensionless parameter}
to motivate a saddle-point approximation.
Hence, the following conceptual question
arises; \emph{where is the classical spacetime?}

Recently, we have suggested to revisit an old idea,
the large-$N$ master field of
Witten~\cite{Witten1979,Coleman1985,GreensiteHalpern1983},
for a possible origin of classical spacetime
in the context of the IIB matrix model~\cite{Klinkhamer2020-master}.
Let us, first, recall the meaning of this mysterious master
field (a name coined by Coleman~\cite{Coleman1985})
and, then, present the final result
(with technical details moved to the Appendices).

\subsection{Large-$N$ master field}
\label{subsec:Large-N-master-field}

Consider the following bosonic observable:
\beq \label{eq:IIB-matrix-model-w-observable}
w^{\mu_{1} \,\ldots\, \mu_{m}} \equiv
\text{Tr}\,\big( A^{\mu_{1}} \cdots\, A^{\mu_{m}}\big)\,,
\eeq
which is invariant under a global gauge transformation,
\beq \label{eq:Amu-prime}
A^{\prime\,\mu}=\Omega\,A^{\,\mu}\,\Omega^{\,\dagger}\,,
\quad
\Omega \in  SU(N) \,.
\eeq
Arbitrary strings of these $w$ observables have expectation values
\beq \label{eq:IIB-matrix-model-w-product-vev}
\hspace*{-0mm}
\langle
w^{\mu_{1}\,\ldots\,\mu_{m}}\:w^{\nu_{1}\,\ldots\,\nu_{n}}\, \cdots\, \rangle
=
\frac{1}{Z}\,\int dA\,
\big(w^{\mu_{1}\,\ldots\,\mu_{m}}\:w^{\nu_{1}\,\ldots\,\nu_{n}}\, \cdots\,\big)
\exp\left[i\,S_\text{eff}/ \ell^{4}\,\right],
\eeq
where the normalization factor $1/Z$ gives $\langle\, 1\, \rangle=1$.

For a string of two identical $w$ observables,
the following factorization property holds to leading order in $N$:
\beq
\label{eq:IIB-matrix-w-square-vev}
\langle \, w^{\mu_{1}\,\ldots\,\mu_{m}}\;w^{\mu_{1}\,\ldots\,\mu_{m}}\,\rangle
\stackrel{N}{=}
\langle w^{\mu_{1}\,\ldots\,\mu_{m}}  \rangle\;
\langle w^{\mu_{1}\,\ldots\,\mu_{m}}  \rangle \,,
\eeq
without sums over repeated indices. 
Similar large-$N$ factorization
properties hold for \emph{all} expectation
values \eqref{eq:IIB-matrix-model-w-product-vev},%
\beq \label{eq:IIB-matrix-model-w-product-vev-factorized}
\hspace*{-3mm}
\langle \,
w^{\mu_{1}\,\ldots\,\mu_{m}}\:w^{\nu_{1}\,\ldots\,\nu_{n}}\, \cdots\,
w^{\omega_{1}\,\ldots\,\omega_{z}}
 \, \rangle
\stackrel{N}{=}
\langle \, w^{\,\mu_{1}\,\ldots\,\mu_{m}} \, \rangle\:
\langle \, w^{\,\nu_{1}\,\ldots\,\nu_{n}} \, \rangle\, \cdots\,
\langle \, w^{\,\omega_{1}\,\ldots\,\omega_{z}} \, \rangle\,,
\eeq
with a product of expectation values on the right-hand side.

The leading-order equality \eqref{eq:IIB-matrix-w-square-vev}
states that the expectation value of the square of $w$ equals
the square of the expectation value of $w$, which
is a truly remarkable result for a  statistical (quantum) theory.
Indeed, according to Witten~\cite{Witten1979}, 
the factorizations \eqref{eq:IIB-matrix-w-square-vev}
and \eqref{eq:IIB-matrix-model-w-product-vev-factorized} imply that
the path integrals \eqref{eq:IIB-matrix-model-w-product-vev}
are \emph{saturated by a single configuration},
the so-called master field $\widehat{A}^{\,\mu}$
(from now on, the caret will denote a master-field variable).

Considering one $w$ observable for simplicity and
working to leading order in $N$, we then have for
its expectation value:%
\beq
\label{eq:IIB-matrix-model-observable-from-master-field}
\langle \, w^{\mu_{1}\,\ldots\, \mu_{m}} \, \rangle
\stackrel{N}{=}
\text{Tr}\,\Big( \widehat{A}^{\,\mu_{1}} \cdots\, \widehat{A}^{\,\mu_{m}}\Big)
\equiv
\widehat{w}^{\,\mu_{1}\,\ldots\, \mu_{m}}  \,.
\eeq
Similarly, we have for the other
expectation values \eqref{eq:IIB-matrix-model-w-product-vev}:%
\beq \label{eq:IIB-matrix-model-w-product-vev-from-master-field}
\hspace*{-3mm}
\langle \,
w^{\mu_{1}\,\ldots\,\mu_{m}}\:w^{\nu_{1}\,\ldots\,\nu_{n}}\, \cdots\,
w^{\omega_{1}\,\ldots\,\omega_{z}}  \, \rangle
\stackrel{N}{=}
\widehat{w}^{\,\mu_{1}\,\ldots\,\mu_{m}}\:
\widehat{w}^{\,\nu_{1}\,\ldots\,\nu_{n}}\, \cdots\,
\widehat{w}^{\,\omega_{1}\,\ldots\,\omega_{z}}\,,
\eeq
with a product of real numbers on the right-hand side.
Hence, we do not have to perform the path integrals
on the right-hand side of \eqref{eq:IIB-matrix-model-w-product-vev}:
we ``only'' need ten traceless Hermitian matrices
$\widehat{A}^{\,\mu}$ to get \emph{all} these expectation values
from the simple procedure of replacing each $A^{\,\mu}$
in the observables by $\widehat{A}^{\,\mu}$
(the quotation marks are there, because the ten matrices are,
of course, very large as $N$ increases towards infinity).

\subsection{Emergent classical spacetime}
\label{subsec:Emergent-classical-spacetime}

Now, the meaning of the suggestion in
the last paragraph of Sec.~\ref{subsec:IIB-matrix-model}
is clear: classical spacetime may reside in the
matrices $\widehat{A}^{\,\mu}$ of the IIB-matrix-model master field.
The heuristics is as follows:
\begin{itemize}
  \item
The expectation values
$\left\langle w^{\mu_{1}\,\ldots\,\mu_{m}}\:w^{\nu_{1}\,\ldots\,\nu_{n}}\,
\cdots\,w^{\omega_{1}\,\ldots\,\omega_{z}} \right\rangle$
from \eqref{eq:IIB-matrix-model-w-product-vev} correspond to
an infinity of real numbers and these real numbers
contain a large part of the \emph{information content}
of the IIB matrix model (but, of course, not all the information).
  \item
That \emph{same} information is carried by
the master-field matrices $\widehat{A}^{\,\mu}$,
which reproduce, to leading order in $N$,
the very same real numbers as the products
$\widehat{w}^{\,\mu_{1}\,\ldots\,\mu_{m}}\,
\widehat{w}^{\nu_{1}\,\ldots\,\nu_{n}} \,\cdots\,
\widehat{w}^{\,\omega_{1}\,\ldots\,\omega_{z}}$,
where each real number $\widehat{w}$ entering
these products is the observable $w$ evaluated for the master-field
ma\-tri\-ces $\widehat{A}^{\,\mu}$.
  \item
From these master-field matrices $\widehat{A}^{\,\mu}$, it appears
indeed feasible to \emph{extract} the points and metric of
an emergent classical spacetime
(recall that the original matrices $A^{\,\mu}$
were merely integration variables).
\end{itemize}
It is certainly satisfying to have a heuristic understanding,
but we would like to proceed further.

Let us, first, \emph{assume} that the matrices $\widehat{A}^{\,\mu}$
of the Lorentzian-IIB-matrix-model master field are known
and that they are approximately band-diagonal
(as suggested by the numerical results of
Refs.~\cite{KimNishimuraTsuchiya2012,NishimuraTsuchiya2019}).
Then, it is possible~\cite{Klinkhamer2020-master}
to extract a discrete set of spacetime points
$\{\widehat{x}^{\,\mu}_{k}\}$
and the emergent inverse metric $g^{\mu\nu}(x)$
for a continuous (interpolating) spacetime coordinate $x^\mu$;
toy-model calculations have been presented in
Refs.~\cite{Klinkhamer2020-points,Klinkhamer2020-metric}.
The metric $g_{\mu\nu}(x)$ is simply obtained as matrix inverse
of $g^{\mu\nu}(x)$.

It is even possible~\cite{Klinkhamer2020-reg-bb-IIB-m-m} that
the large-$N$ master field of the Lorentzian IIB matrix model
gives the regularized-big-bang metric \eqref{eq:reg-bb}.
In that case, the final result is that the effective length parameter
$b$ of the emergent regularized-big-bang
metric \eqref{eq:reg-bb} 
is calculated in terms of the IIB-matrix-model length scale $\ell$:%
\beq
b_\text{eff} \sim               \ell
             \stackrel{?}{\sim} l_\text{Planck}
             \equiv  \sqrt{\hbar\,G/c^3}
             \approx 1.62 \times 10^{-35}\,\text{m}  \,.
\eeq
The argument for the connection of $\ell$
and $l_\text{Planck}$ has been presented in Sec.~IV
of Ref.~\cite{Klinkhamer2020-reg-bb-IIB-m-m},
under the assumption that Einstein gravity
is recovered from the IIB matrix model.

Technical details are collected in four Appendices.
Specifically,
Appen\-dix~\ref{app:Extraction-spacetime-points}
discusses how precisely the spacetime  points
are extracted from the matrices $\widehat{A}^{\,\mu}$
of the Lorentzian-IIB-matrix-model master field,
under the assumption that these matrices are known.
Appendix~\ref{app:Extraction-spacetime-metric}
then shows how the emergent spacetime metric
is obtained from the distributions of the extracted spacetime points.
Appendix~\ref{app:Various-emergent-spacetimes}
discusses different emergent spacetimes, which result from
different assumptions for the properties of the master-field matrices.
(Discussed as well, in the second subsection of 
Appendix~\ref{subapp:Emergent-Minkowski-and-RW-metrics}, 
is the issue of ``topology without topology,''
namely obtaining an effective nontrivial topology
from strong gravitational fields~\cite{Klinkhamer2012}.)
Appendix~\ref{app:More-on-the-Lorentzian-signature}
presents an alternative mechanism for getting
a Lorentzian signature of the emergent spacetime metric.

At this moment, we should  mention that there are also
other approaches to extracting a classical spacetime,
examples being noncommutative geometry~\cite{Connes2000,Steinacker2019}
and entanglement~\cite{vanRaamsdonk2020,Das-etal2020a}.
\vspace*{-1mm}  

\vspace*{-1mm}
\section{Conclusion}
\label{sec:Conclusion}

It is conceivable that a new physics phase replaces
the Friedmann big bang singularity suggested by our current
theories, general relativity and the standard model of elementary
particles. For an explicit calculation, we have
turned to the IIB matrix model, which has been proposed as
a nonperturbative formulation of type-IIB superstring
theory (M-theory).

The crucial insight is that the emergent classical spacetime may
reside in the matrices $\widehat{A}^{\,\mu}$ of the
large-$N$ master field of the IIB matrix model. The
master-field  matrices $\widehat{A}^{\,\mu}$
can, in principle, produce the regularized-big-bang metric
\eqref{eq:reg-bb} with length parameter $b\sim \ell$,
where $\ell$ is the length scale of the matrix model.

The outstanding task, now, is to \emph{calculate} the exact
matrices $\widehat{A}^{\,\mu}$ of the IIB-matrix-model master field
(see also the last subsection of 
Appendix~\ref{app:More-on-the-Lorentzian-signature} 
for further comments). 
As the exact matrices $\widehat{A}^{\,\mu}$ will be hard to obtain,
it perhaps makes sense to first look
for a reliable approximation of them.
\vspace*{-3mm}  

\section*{\hspace*{-4.5mm}Acknowledgments}  
\noindent
It is a pleasure to thank Z.L. Wang for discussions
on cosmological singularity theorems
and the Conference Organizers for bringing about 
this interesting online meeting.  
\vspace*{-4mm}  

\section*{\hspace*{-4.5mm}Note Added}  
\noindent
Following up on the remarks in the very last paragraph
of Appendix~\ref{app:More-on-the-Lorentzian-signature},
we have started with the calculation of the
master-field matrices~\cite{Klinkhamer2021-first-look,%
Klinkhamer2021-solutions-exact}.
\vspace*{-2mm}

\begin{appendix}
\section{Extraction of the spacetime points}
\label{app:Extraction-spacetime-points}

Aoki et al.~\cite{Aoki-etal-review-1999}
have argued that the \emph{eigenvalues} of the bosonic matrices
$A^{\,\mu}$ of model \eqref{eq:IIB-matrix-model-ZwithSgeneral-S-eta}
can be interpreted as \emph{spacetime coordinates}, so that the model
has a ten-dimensional $\mathcal{N}=2$ spacetime supersymmetry.
This supersymmetry, incidentally, implies the existence of a graviton,
as long as there are massless particles in the spectrum.

Here, we will turn to the eigenvalues
of the \emph{master-field} matrices $\widehat{A}^{\,\mu}$.
Assume that the matrices $\widehat{A}^{\,\mu}$
of the Lorentzian-IIB-matrix-model master field are known and that
they are approximately band-diagonal 
with width $\Delta N < N$  (as suggested by the numerical results of
Refs.~\cite{KimNishimuraTsuchiya2012,NishimuraTsuchiya2019}).
Then, make a particular global gauge
transformation~\cite{KimNishimuraTsuchiya2012},%
\beqa \label{eq:Amuhat-bar}
\underline{\widehat{A}}^{\,\mu}
&=&
\underline{\Omega}\,\widehat{A}^{\,\mu}\,\underline{\Omega}^{\,\dagger}\,,
\quad
\underline{\Omega} \in  SU(N) \,,
\eeqa
so that the transformed 0-component matrix is diagonal
(the component 0 is singled out by
the ``Lorentzian'' coupling constants $\widetilde{\eta}_{\mu\nu}$)
and has ordered 
eigenvalues $\widehat{\alpha}_{i} \in \mathbb{R}$,%
\bsubeqs\label{eq:A0hat-bar-alphahat-order}
\beqa \label{eq:A0hat-bar}
\underline{\widehat{A}}^{\,0}
&=&
\text{diag} \big( \widehat{\alpha}_{1},\,\widehat{\alpha}_{2},
\,\ldots\,,\,
\widehat{\alpha}_{N-1},\,\widehat{\alpha}_{N} \big)\,,
\\[1mm]
\label{eq:alphahat-order}
\widehat{\alpha}_{1}
&\leq&
\widehat{\alpha}_{2}\,\leq\,\;\ldots\;\,\leq\,
\widehat{\alpha}_{N-1}\,\leq\,\widehat{\alpha}_{N}\,,
\\[1mm]
\sum_{i=1}^{N}\,\widehat{\alpha}_{i} &=& 0\,.
\eeqa
\esubeqs
The ordering \eqref{eq:alphahat-order} will turn out to be crucial for
the time coordinate $\widehat{t}\,$  to be obtained later.

A relatively simple procedure~\cite{Klinkhamer2020-master}
\emph{approximates} the eigenvalues of the
spatial
matrices $\underline{\widehat{A}}^{\,m}$ but still manages
to \emph{order them along the diagonal},
in sync with the temporal eigenvalues $\widehat{\alpha}_{i}$ from
\eqref{eq:A0hat-bar-alphahat-order}.
This procedure corresponds, in fact,
to a type of coarse graining of some of the
information contained in the master field.

We start from the following trivial observations.
If $M$ is an $N\times N$ Hermitian matrix, then any
$n\times n$ block centered on the diagonal of $M$
is also Hermitian, which holds for any value of $n$
with $1 \leq n \leq N$. If the matrix $M$ is, moreover,
band-diagonal with width $ \Delta N < N$, then the
eigenvalues of the $n\times n$ blocks on the diagonal
approximate the original eigenvalues of $M$,
provided $n \gtrsim \Delta N$.

Now, let $K$ and $n$ be divisors of $N$, so that
\beqa\label{eq:N-as-product-K-times-n}
N&=&K\,n\,,
\eeqa
where both $K$ and $n$ are positive integers.
Then, consider, in each of the ten matrices $\underline{\widehat{A}}^{\,\mu}$,
the $K$ adjacent blocks of size $n\times n$ centered on the diagonal.

From \eqref{eq:A0hat-bar-alphahat-order},
we already know the diagonalized $n\times n$ blocks of  
$\underline{\widehat{A}}^{\,0}$
with eigenvalues $\widehat{\alpha}_{i}$.
This allows us to define the following time coordinate
$\widehat{t}\,(\sigma)$ for $\sigma\in (0,\,1]$
from the averages of the $\widehat{\alpha}_{i}$'s
in the blocks (labeled $k$):  
\beq \label{eq:x0hat-def}
\widehat{x}^{\,0}\,\big(k/K\big) \equiv
\widetilde{c}\;\widehat{t}\,\big(k/K\big) =
\frac{1}{n}\;\sum_{j=1}^{n} \, \widehat{\alpha}_{(k-1)\,n+j}\,,
\eeq
with  $k \in \{1,\,\ldots ,\,  K\}$ and a
velocity $\widetilde{c}$ to be set to unity later.
The time coordinates from \eqref{eq:x0hat-def} are ordered,
\beq
\label{eq:that-order}
\widehat{t}\,\big(1/K\big) \,\leq\, \widehat{t}\,\big(2/K\big)\,\leq
\;\ldots\;
\leq\, \widehat{t}\,\big(1-1/K\big)\,\leq\, \widehat{t}\,\big(1\big)\,,
\eeq
precisely because the $\widehat{\alpha}_{i}$ 
are ordered according to \eqref{eq:alphahat-order}.  
Observe that this ordering property
is the defining characteristic of what makes a physical time.

Next, obtain the eigenvalues of the $n\times n$ blocks of the nine
spatial matrices $\underline{\widehat{A}}^{\,m}$
and denote these real eigenvalues by
$\big(\widehat{\beta}^{\,m}\big)_{i}\,$,
with a label \mbox{$i \in \{1,\,\ldots ,\,  N\}$}
respecting the order of the $n$-dimensional blocks.
Define, just as for the time coordinate in \eqref{eq:x0hat-def},
the following nine spatial coordinates $\widehat{x}^{\,m}(\sigma)$
for $\sigma \in (0,\,1]$ from the respective averages:
\beq \label{eq:xmhat-def}
\widehat{x}^{\,m}\big(k/K\big)
=
\frac{1}{n}\;\sum_{j=1}^{n}\, \left[\,\widehat{\beta}^{\,m}\,\right]_{(k-1)\,n+j}\,,
\eeq
with  $k \in \{1,\,\ldots ,\,  K\}$.

If the master-field matrices  $\underline{\widehat{A}}^{\,\mu}$
are approximately \emph{band-diagonal} with width $\Delta N$
and if the eigenvalues of the spatial $n\times n$ blocks
(with $n \gtrsim \Delta N$)
show significant \emph{scattering}~\cite{Klinkhamer2020-points},
then the expressions \eqref{eq:x0hat-def} and \eqref{eq:xmhat-def}
may provide suitable spacetime points.
In a somewhat different notation, these points are denoted
\beq
\label{eq:xhat-mu-k}
\widehat{x}^{\,\mu}_{k}=
\left(\,\widehat{x}^{\,0}_{k},\, \widehat{x}^{\,m}_{k}\,\right)
\equiv
\Big(\,\widehat{x}^{\,0}\big(k/K\big),\,
\widehat{x}^{\,m}\big(k/K\big)\,\Big)\,,
\eeq
where $k$ runs over $\{1,\,\ldots ,\,  K\}$
with $K=N/n$ from \eqref{eq:N-as-product-K-times-n}.
Each of these coordinates $\widehat{x}^{\,\mu}_{k}$
has the dimension of length, which
traces back to the dimension of the bosonic matrix
variable $A^{\mu}$, as mentioned 
in the second technical remark of the new paragraph
starting a few lines below \eqref{eq:IIB-matrix-model-ZwithSgeneral-S-eta}.  

There are alternative procedures for the
extraction of spacetime points, one of which is discussed in
Appendix~B of Ref.~\cite{Klinkhamer2020-points}.
That alternative procedure randomly selects
\emph{one} eigenvalue from \emph{each} $n\times n$ block
of the gauge-transformed
master-field matrices  $\underline{\widehat{A}}^{\,\mu}$ and gives
the following extracted points (denoted by a tilde
instead of a caret):
\bsubeqs\label{eq:xtilde-mu-k}
\beqa
\widetilde{x}^{\,0}_{k} &=&
\widehat{\alpha}_{(k-1)\,n+\text{rand}[1,\,n]}\,,
\\[1.0mm]
\widetilde{x}^{\,m}_{k} &=&
\left[\,\widehat{\beta}^{\,m}\,\right]_{(k-1)\,n+\text{rand}[1,\,n]}\,,
\eeqa
\esubeqs
where $k$ runs over $\{1,\,\ldots ,\,  K\}$ and
``$\text{rand}[1,\,n]$'' is a uniform pseudorandom integer from the
set \mbox{$\{1,\, 2,\,\ldots\,,\, n\}$.}
For the moment, we focus our attention on the averaging
procedure from \eqref{eq:x0hat-def} and  \eqref{eq:xmhat-def}.

To summarize, with $N=K\,n$  and $n \gtrsim \Delta N$,
the extracted spacetime points $\widehat{x}^{\,\mu}_{k}$,
for $k \in \{1,\,\ldots ,\,  K\}$, are obtained as
\emph{averaged eigenvalues} of
the $n\times n$ blocks along the diagonals of the gauge-transformed
master-field matrices  $\underline{\widehat{A}}^{\,\mu}$
from \eqref{eq:Amuhat-bar}--\eqref{eq:A0hat-bar-alphahat-order}.

\vspace*{-2mm}
\section{Extraction of the spacetime metric}
\label{app:Extraction-spacetime-metric}

The discrete set of points $\{\widehat{x}^{\,\mu}_{k}\}$
from \eqref{eq:xhat-mu-k}
effectively builds a spacetime manifold with continuous
(interpolating) coordinates $x^{\mu}$
if there is also an emerging metric $g_{\mu\nu}(x)$.

From the effective action of a low-energy
scalar degree of freedom $\sigma$ ``propagating'' over the discrete
spacetime points $\widehat{x}^{\,\mu}_{k}$
(see below for details), the
following expression for the emergent \emph{inverse} metric
has been obtained~\cite{Aoki-etal-review-1999,%
Klinkhamer2020-master}:%
\bsubeqs\label{eq:emergent-inverse-metric-gMUNU-rho-av}
\beqa \label{eq:emergent-inverse-metric-gMUNU}
\hspace*{-4mm}
g^{\mu\nu}(x) &\sim&
\int_{\mathbb{R}^{D}} d^{D}y\;
\rho_\text{av}(y)\; (x-y)^{\mu}\,(x-y)^{\nu}\;f(x-y)\;r(x,\,y)\,,
\\[1.0mm] 
\label{eq:emergent-inverse-metric-rho-av}
\hspace*{-4mm}
\rho_\text{av}(y) &\equiv&
\langle\langle\, \rho(y)  \,\rangle\rangle\,,
\eeqa
\esubeqs
with continuous spacetime coordinates $x^\mu$ having the
dimension of length and spacetime dimension $D=1+9=10$ for
the original model \eqref{eq:IIB-matrix-model-ZwithSgeneral-S-eta}.

The quantities that enter the 
integral \eqref{eq:emergent-inverse-metric-gMUNU-rho-av}
are the density function
\vspace*{0mm}
\beq \label{eq:rho-def}
\rho(x) \;\equiv \;
\sum_{k=1}^{K}\;\delta^{(D)} \big(x- \widehat{x}_{k}\big)\,,
\vspace*{0mm}
\eeq
the density correlation function $r(x,\,y)$ defined by
\vspace*{0mm}
\beq \label{eq:r-def}
\langle\langle\,\rho(x)\,\rho(y) \,\rangle\rangle \;\equiv \;
\langle\langle\, \rho(x)\,\rangle\rangle\; \langle\langle\,\rho(y) \,\rangle\rangle\; r(x,\,y)\,,
\vspace*{0mm}
\eeq
and a localized real function $f(x)$
from the scalar effective action,
\vspace*{-1mm}
\begin{equation} \label{eq:Seff-sigma}
S_\text{eff}[\sigma] \sim
\sum_{k,\,l}\; \frac{1}{2}\,f\big(\widehat{x}_{k}-\widehat{x}_{l}\big)\;
\big( \sigma_{k}- \sigma_{l}  \big)^{2}\,,
\vspace*{-2mm}
\end{equation}
where $\sigma_{k}$ is the real field value
at the point $\widehat{x}_{k}$
(the real scalar degree of freedom $\sigma$
arises from a perturbation of the
master field $\underline{\widehat{A}}^{\,\mu}$
and has the dimension of length;
see Appendix~A in Ref.~\cite{Klinkhamer2020-master}
for a toy-model calculation).
For the extraction procedure of
Appendix~\ref{app:Extraction-spacetime-points},
the quantity $\langle\langle\, \rho(y)  \,\rangle\rangle$
in \eqref{eq:emergent-inverse-metric-rho-av}
results from averaging over different block sizes $n$
and block positions along the diagonals in the master-field
matrices $\underline{\widehat{A}}^{\,\mu}$
(possibly with smaller blocks at the end or beginning of the
diagonal, for a fixed large value of $N$).

Very briefly,
expression \eqref{eq:emergent-inverse-metric-gMUNU-rho-av}
is obtained as follows from the effective action \eqref{eq:Seff-sigma}.
Define the continuous field $\sigma(x)$
as having $\sigma(\widehat{x}_{k})=\sigma_{k}\,$ and
write \eqref{eq:Seff-sigma} in terms of  $\sigma(x)$.
Next, average over different block structures
in the master-field matrices (see above)
and make appropriate Taylor expansions of $\sigma(x)$.
The continuous field $\sigma(x)$ is then found to have
a standard local kinetic term
$(1/2)\,\partial_{\mu} \sigma(x)\,\partial_{\nu}\sigma(x)\,g^{\mu\nu}(x)$
in the action, with the inverse metric $g^{\mu\nu}(x)$ as given
by the expression \eqref{eq:emergent-inverse-metric-gMUNU-rho-av}.
See Sec.~4.2 of
Ref.~\cite{Aoki-etal-review-1999} for further details.

As $r(x,\,y)$ is dimensionless and
$f(x)$ has dimension $1/(\text{length})^{2}$,
the inverse metric $g^{\mu\nu}(x)$
from \eqref{eq:emergent-inverse-metric-gMUNU-rho-av} is seen
to be dimensionless.
The metric $g_{\mu\nu}$ is obtained as the matrix inverse
of $g^{\mu\nu}$.
See Sec. II B of Ref.~\cite{Klinkhamer2020-reg-bb-IIB-m-m}
for some  heuristic remarks on expression
\eqref{eq:emergent-inverse-metric-gMUNU-rho-av}
for the emergent inverse metric.

Note that, in principle, the origin of 
expression \eqref{eq:emergent-inverse-metric-gMUNU-rho-av}
need not be the IIB matrix model but can be an entirely
different theory, as long as the emerging inverse metric
is given by a multiple integral with the same basic structure.
But, for the moment, we only discuss a IIB-matrix-model origin.

To summarize, the emergent metric, in the context of the IIB matrix model,
is obtained from \emph{correlations} of the extracted spacetime points and
the master-field perturbations.

\vspace*{-2mm}
\section{Various emergent spacetimes}
\label{app:Various-emergent-spacetimes}

\vspace*{-1.5\baselineskip}  
\subsection*{Preliminaries} 

The obvious question, now, is which spacetime and
metric \emph{do} we get from the steps outlined
in Appendices~\ref{app:Extraction-spacetime-points}
and \ref{app:Extraction-spacetime-metric}.
We don't know for sure, as we do not have the exact
IIB-matrix-model master field.
But, awaiting the final result on the master field, we can already
investigate what properties the master field \emph{would}
need to have in order to be able to produce certain desired metrics.
It is far from obvious that these desired metrics can be obtained from
expression \eqref{eq:emergent-inverse-metric-gMUNU-rho-av},
but it appears indeed feasible.
First results are presented in the rest of this appendix.

\subsection*{Emergent Minkowski and Robertson--Walker metrics}

We restrict ourselves to four ``large'' spacetime
dimensions~\cite{KimNishimuraTsuchiya2012,NishimuraTsuchiya2019},
setting
\beq \label{eq:D-equals-4}
D=1+3=4\,,
\eeq
and use length units that normalize 
the Lorentzian-IIB-matrix-model length scale,%
\beq \label{eq:ell-equals-1}
\ell=1\,.
\eeq
In Ref.~\cite{Klinkhamer2020-metric}, we have then shown that
it is possible to choose appropriate functions
$\rho_\text{av}(y)$, $f(x-y)$, and $r(x,\,y)$
in \eqref{eq:emergent-inverse-metric-gMUNU-rho-av} for $D=4$,
so that the  Minkowski metric is obtained,
as given by \eqref{eq:RW-ds2} for $a^{2}(t)=1$.
Furthermore, it is possible to deform the chosen function
$\rho_\text{av}(y)$, so that the spatially flat ($k=0$)
Robertson--Walker metric \eqref{eq:RW-ds2} is obtained.

The question was raised in Ref.~\cite{Klinkhamer2020-metric}
whether or not it would also be possible, in principle, to
obtain a Robertson--Walker universe with positive ($k=+1$)
or negative ($k=-1$) spatial curvature.
This is perhaps not excluded as the following result suggests.
It has been shown, in fact, that the $k=1$ Robertson--Walker universe
need not really have an underlying $\mathbb{R}\times S^{3}$ topology
but can result from strong gravitational fields over Minkowski
spacetime with an $\mathbb{R}^{4}$ topology
(see Ref.~\cite{Klinkhamer2012}, which builds upon
unpublished work from 1992 by M. Veltman and the present author).
%

\subsection*{Emergent regularized-big-bang metric}

In order to get an inverse metric whose component $g^{00}$
\emph{diverges} at $t=0$, it is necessary to \emph{relax} the
convergence properties of the $y^0$
integral in \eqref{eq:emergent-inverse-metric-gMUNU-rho-av}
by a suitable change in the \textit{Ans\"{a}tze} for
$\rho_\text{av}(y)$, $f(x-y)$, and $r(x,\,y)$.

In this way, we are able to obtain the following
inverse metric~\cite{Klinkhamer2020-reg-bb-IIB-m-m}:
\beq
\label{eq:gMUNU-emerged-inverse-metric}
\hspace*{-0mm}
g^{\mu\nu}_\text{(eff)}
 \sim
\begin{cases}
     -\ddfracNEW{\textstyle t^{2}+c_{-2}}{\textstyle t^{2}}
 \,,   &  \;\;\text{for}\;\;\mu=\nu=0 \,,
 \\[3mm]
 1 + c_{2}\,t^{2} + c_{4}\,t^{4} + \dots  \,,
 &  \;\;\text{for}\;\;\mu=\nu=m\in \{1,\, 2,\, 3\} \,,
 \\[3mm]
 0 \,,   &  \;\;\text{otherwise}\,,
 \end{cases}
\eeq
with real dimensionless coefficients $c_{n}$ in $g^{mm}_\text{(eff)}$ 
resulting from the requirement that $t^{n}$ terms, for $n>0$, vanish
in $g^{00}_\text{(eff)}$.
The matrix inverse of \eqref{eq:gMUNU-emerged-inverse-metric}
gives the following Lorentzian metric:
\beq
\label{eq:gmunu-emerged-metric}
\hspace*{-0mm}
g_{\mu\nu}^\text{(eff)}
\sim
\begin{cases}
  - \ddfracNEW{\textstyle t^{2}}{\textstyle t^{2}+c_{-2}}
 \,,   &  \;\;\text{for}\;\;\mu=\nu=0 \,,
 \\[3mm]
 \ddfracNEW{\textstyle  1}{\textstyle 1 + c_{2}\,t^{2} + c_{4}\,t^{4} + \dots } \;,
 &  \;\;\text{for}\;\;\mu=\nu=m\in \{1,\, 2,\, 3\} \,,
 \\[3mm]
 0 \,,   &  \;\;\text{otherwise}\,,
 \end{cases}
\eeq
which has, for $c_{-2}> 0$, a vanishing determinant at $t=0$
and is, therefore, degenerate.

The emergent degenerate metric \eqref{eq:gmunu-emerged-metric}
has, in fact,
the structure of the RWK metric \eqref{eq:reg-bb},
with the following effective parameters:
\beq\label{eq:b2-eff-a-eff}
b^{2}_\text{eff} \sim c_{-2}\;\ell^{2}\,,
\qquad
a^{2}_\text{eff}(t) \sim 1-c_{2}\;\big(t/\ell\big)^{2} + \ldots\,,
\eeq
where the IIB-matrix-model length scale $\ell$ has been
restored and where the leading coefficients
$c_{-2}$ and $c_{2}$ have been
calculated~\cite{Klinkhamer2020-reg-bb-IIB-m-m}.
By choosing the \textit{Ansatz} parameters appropriately,
we can get $c_{-2}>0$ and $c_{2} <0$ in \eqref{eq:b2-eff-a-eff},
so that the emergent classical spacetime corresponds to
the spacetime of a nonsingular cosmic bounce at $t=0$,
as obtained in \eqref{eq:regularized-Friedmann-asol}
from Einstein's gravitational field equation
with a homogeneous relativistic perfect fluid.
Incidentally, a possibly odd functional behavior of $a(t)$, as mentioned
in point 2 of Sec.~\ref{subsec:Bounce-or-new-physics-phase?},
could result from a consistency condition involving the
fermionic degrees of freedom, which are present in
the original matrix model \eqref{eq:IIB-matrix-model-ZwithSgeneral-S-eta}. 

As a final remark, we note that the \emph{origin}
of the spacetime defect in the regularized-big-bang (RWK) metric
appears to be due
to \emph{long-range tails}~\cite{Klinkhamer2020-reg-bb-IIB-m-m}
of certain correlation functions entering the multiple-integral
expression  \eqref{eq:emergent-inverse-metric-gMUNU-rho-av}
for the emergent inverse metric.

\section{More on the Lorentzian signature}
\label{app:More-on-the-Lorentzian-signature}

\vspace*{-1.5\baselineskip}  
\subsection*{Alternative mechanism}

Up till now, we have considered the Lorentzian IIB matrix model,
which has two characteristics:
\begin{enumerate}
\item
the Feynman phase factor $\exp\left[i\,S/ \ell^{4}\,\right]$
in the ``path'' integral
\eqref{eq:IIB-matrix-model-ZwithSgeneral};
\item
the ``Lorentzian'' coupling constants $\widetilde{\eta}_{\mu\nu}$
from \eqref{eq:IIB-matrix-model-etamunu} entering
the action \eqref{eq:IIB-matrix-model-action}.
\end{enumerate}
With an assumed master field of this Lorentzian matrix model,
we obtained the spacetime points from expressions \eqref{eq:x0hat-def}
and \eqref{eq:xmhat-def} in Appendix~\ref{app:Extraction-spacetime-points} and
the inverse metric from expression
\eqref{eq:emergent-inverse-metric-gMUNU-rho-av}
in Appendix~\ref{app:Extraction-spacetime-metric}.

Several Lorentzian inverse metrics were found in
Appendix~\ref{app:Various-emergent-spacetimes}, where the used
\textit{Ans\"{a}tze}~\cite{Klinkhamer2020-metric,Klinkhamer2020-reg-bb-IIB-m-m}
relied on the available ``Lorentzian'' coupling
constants $\widetilde{\eta}_{\mu\nu}$.
Specifically, the metrics obtained were
the Minkowski metric, the Robertson--Walker metric,
and the regularized-big-bang (RWK) metric.

There is, however, another way~\cite{Klinkhamer2020-master} 
to obtain Lorentzian inverse metrics,
namely by making an appropriately \emph{odd}
\textit{Ansatz} for the correlations functions entering
expression \eqref{eq:emergent-inverse-metric-gMUNU-rho-av},
so that the resulting matrix has an \emph{off-diagonal}
structure with one real eigenvalue having a different sign
than the others.

With this appropriately odd \textit{Ansatz},
it is, in principle, also possible
to get a \emph{Lorentzian} emergent inverse metric from
the \emph{Euclidean} matrix model, which has
a weight factor $\exp\left[-S/ \ell^{4}\,\right]$ in the path integral
and nonnegative coupling constants $\widetilde{\delta}_{\mu\nu}$
in the action.
The spacetime points are extracted from the Euclidean master field
(without need for a particular gauge transformation) by 
expression \eqref{eq:xmhat-def},
where $m$ now runs over $\{1,\,\ldots ,\,  D\}$.

\subsection*{Toy-model calculation}

We present, here, the full details of a Euclidean toy-model calculation,
which appeared as a parenthetical remark in the last paragraph
of Appendix~B in Ref.~\cite{Klinkhamer2020-master}.

We start the calculation with the
multiple integral \eqref{eq:emergent-inverse-metric-gMUNU-rho-av},
for spacetime dimension $D=4$ and model length scale $\ell=1$,
and write in the integrand%
\beq\label{eq:h-rbar-del}
f(x-y)\;r(x,\,y)= f(x-y)\;\widetilde{r}(y-x)\;\overline{r}(x,\,y)
= h(y-x)\;\overline{r}(x,\,y)\,,
\eeq
where the new function $\overline{r}(x,\,y)$
has a more complicated dependence on $x$ and $y$ 
than the combination $x-y$.  

The $D=4$
multiple integral \eqref{eq:emergent-inverse-metric-gMUNU-rho-av},
having $y^{0}$ replaced by $y^{4}$,
is then evaluated at the spacetime point
\begin{subequations}\label{eq:appD-assumptions}
\begin{equation} \label{eq:appD-assumptions-xmu-equals-zero}
x^{\mu} = 0 \,,
 \end{equation}
with the replacement \eqref{eq:h-rbar-del} in the integrand,
two further simplifications,
\begin{equation} \label{eq:appD-assumptions-rho-av-equals-1-rbar-equals-1}
\rho_\text{av}(y) = 1\,,
\qquad
\overline{r}(x,\,y)= 1\,,
\end{equation}
and symmetric cutoffs on the integrals,
\begin{eqnarray} \label{eq:appD-assumptions-symmetric-cutoffs}
\int_{-1}^{1} dy^{1}\, \ldots \,\int_{-1}^{1} dy^{4}\,.
\end{eqnarray}
\end{subequations}
The only nontrivial contribution to the integrand
of \eqref{eq:emergent-inverse-metric-gMUNU-rho-av}
then comes from the correlation function $h$ as defined by
\eqref{eq:h-rbar-del}.

From  \eqref{eq:emergent-inverse-metric-gMUNU-rho-av},
\eqref{eq:h-rbar-del}, and \eqref{eq:appD-assumptions},
we get the following expression for
the emergent inverse metric at $x^\mu=0$:
\begin{equation}
\label{eq:inv-metric-testE4-integrals}
g^{\mu\nu}_\text{test,E4}(0) \sim
\int_{-1}^{1} dy^{1} \int_{-1}^{1} dy^{2} \int_{-1}^{1} dy^{3} \int_{-1}^{1} dy^{4}
\;\; y^{\mu}\,y^{\nu}\;h(y)\,.
\end{equation}
Next, make an appropriate \textit{Ansatz} 
for the correlation
function $h$ in \eqref{eq:inv-metric-testE4-integrals}:%
\beq\label{eq:Ansatz-htestE4}
h(y)=  1 - \gamma\,\big(y^{1}\, y^{2} + y^{1}\, y^{3}
+ y^{1}\, y^{4} + y^{2}\, y^{3} + y^{2}\, y^{4} + y^{3}\, y^{4}\big)\,,
\eeq
where $\gamma$ multiplies monomials that are odd in two coordinates
and even in the two others.
Note that the \textit{Ansatz} \eqref{eq:Ansatz-htestE4}
treats \emph{all} coordinates $y^{1}$,  $y^{2}$, $y^{3}$,
and $y^{4}$ \emph{equally}, matching the structure of the
coupling constants $\widetilde{\delta}_{\mu\nu}$
of the Euclidean matrix model.

The integrals of \eqref{eq:inv-metric-testE4-integrals}
with \textit{Ansatz} function
\eqref{eq:Ansatz-htestE4} are trivial and we obtain
\begin{subequations}\label{eq:inv-metric-testE4-result-eigenval}
\begin{eqnarray} \label{eq:inv-metric-testE4-result}
\hspace*{-10mm}
g^{\mu\nu}_{\gamma}(0)
&\sim&
\frac{16}{9}\,
\left(
  \begin{array}{cccc}
3 & \;-\gamma\; & \;-\gamma\; & \;-\gamma\;\\
\;-\gamma\; & 3 & -\gamma & -\gamma\\
-\gamma & -\gamma &  3 &  -\gamma\\
-\gamma & -\gamma & -\gamma &3\\
  \end{array}
\right),
\end{eqnarray}
where the matrix on the right-hand side
has the following eigenvalues and corresponding
(normalized) eigenvectors:%
\begin{eqnarray} \label{eq:inv-metric-testE4-eigenval}
\hspace*{-9mm}
\mathcal{E}_{\gamma} &=&
\frac{16}{9}\,\Big\{
\left( 3-3\,\gamma \right),\,
\left( 3+\gamma \right),\,
\left( 3+\gamma \right),\,
\left( 3+\gamma \right)
\Big\}\,,
\\[1mm]
\label{eq:inv-metric-testE4-eigenvectors}
\hspace*{-9mm}
\mathcal{V}_{\gamma} &=&
\left\{
\frac{1}{2}\left(  \begin{array}{c}
    1 \\
    1 \\
    1 \\
    1 \\
  \end{array}\right),\,
\frac{1}{\sqrt{2}}\left(  \begin{array}{c}
    1 \\
    -1 \\
    0 \\
    0 \\
  \end{array}\right),\,
\frac{1}{\sqrt{2}}\left(  \begin{array}{c}
    0 \\
    0 \\
    1 \\
    -1 \\
  \end{array}\right),\,
\frac{1}{2}\left(  \begin{array}{c}
    1 \\
    1 \\
    -1 \\
    -1 \\
  \end{array}\right)
\right\}.
\end{eqnarray}
\end{subequations}

From the eigenvalues \eqref{eq:inv-metric-testE4-eigenval}, we get
the following signatures:
\begin{subequations}\label{eq:inv-metric-testE4-signatures}
\begin{eqnarray}
(+ - - - )\,,\quad  &\text{for}& \gamma \in (-\infty ,\, -3) \,,
\\[1mm]
(+ + + + )\,,\quad  &\text{for}& \gamma \in (-3,\, 1)  \,,
\\[1mm]
(- + + + )\,,\quad  &\text{for}& \gamma \in (1,\,\infty)  \,.
\end{eqnarray}
\end{subequations}
Hence, we find Lorentzian signatures
for parameter values $\gamma$ sufficiently far away from zero,
$\gamma>1$ or $\gamma<-3$.
Note that, for the Lorentzian cases, the time direction
corresponds to the first eigenvector
of \eqref{eq:inv-metric-testE4-eigenvectors} with four equal components.
%

\vspace*{-3mm}
\subsection*{Outlook}
\vspace*{-0mm}

The above toy-model calculation has shown 
that it is, in principle, possible to get a Lorentzian emergent
inverse metric from the Euclidean IIB matrix model,
provided the correlation functions have an appropriate structure.
This observation, if applicable,  would remove the
need for working with the possibly more difficult Lorentzian IIB matrix
model~\cite{KimNishimuraTsuchiya2012,NishimuraTsuchiya2019}
and we could return to the original Euclidean IIB matrix
model~\cite{IKKT-1997,Aoki-etal-review-1999}.

In this respect, it is worth mentioning that, from earlier work
by Greensite and Halpern~\cite{GreensiteHalpern1983},
we have obtained~\cite{Klinkhamer2020-master}
an \emph{algebraic equation} for the Euclidean bosonic master field.
It remains to solve this equation $\ldots$

\end{appendix}

\newpage  


\begin{thebibliography}{99}
\vspace*{-2mm}


\bibitem{Einstein1916}
A.~Einstein,
\hspace*{0.00mm}``Die Grundlage der allgemeinen Relativit\"{a}tstheorie''
\hspace*{0.00mm}(The foundation of the general theory of relativity),
\textit{Annalen Phys. (Leipzig)} {\bf 49}, 769 (1916).

\bibitem{Friedmann1922-1924}
A.A. Friedmann,
\hspace*{0.00mm}``\"{U}ber die Kr\"{u}mmung des Raumes''
\hspace*{0.00mm} (On the curvature of space),
\textit{Z. Phys.} {\bf 10}, 377 (1922);
\hspace*{0.00mm}``\"{U}ber die M\"{o}glichkeit einer Welt mit konstanter negativer
\hspace*{0.00mm}Kr\"{u}mmung des Raumes''
\hspace*{0.00mm}(On the possibility of a world with constant negative curvature),
\hspace*{0.00mm}\textit{Z. Phys.}
{\bf 21}, 326 (1924).

\bibitem{Lemaitre1927}
G. Lema\^{i}tre,
\hspace*{0.00mm}``Un univers homog\`{e}ne de masse constante et de rayon croissant
\hspace*{0.00mm}rendant compte de la vitesse radiale des n\'{e}buleuses extra-galactiques''
\hspace*{0.00mm}(A homogeneous universe of constant mass and increasing radius
\hspace*{0.00mm}accounting for the radial velocity of extra-galactic nebulae),
\textit{Ann. Soc. Sci. Bruxelles} \textbf{A47}, 49 (1927).
%
%
%

\bibitem{Robertson1935}
H.P. Robertson,
\hspace*{0.00mm}``Kinematics and world-structure,'
\textit{Astrophys. J.} \textbf{82}, 284 (1935);
\textbf{83}, 187 (1936); \textbf{83}, 257 (1936).
%


\bibitem{Walker1937}
A.G. Walker, 
\hspace*{0.00mm}``On Milne's theory of world-structure,''
\textit{Proc. Lond. Math. Soc.} \textbf{42}, 90 (1937).
%




\bibitem{Klinkhamer2019-rbb}
F.R.~Klinkhamer,
\hspace*{0.00mm}``Regularized big bang singularity,''
\textit{Phys. Rev. D} \textbf{100}, 023536 (2019)
[arXiv:1903.10450 [gr-qc]].



\bibitem{Klinkhamer2020-more}
F.R.~Klinkhamer,
\hspace*{0.00mm}``More on the regularized big bang singularity,''
\textit{Phys. Rev. D} \textbf{101}, 064029 (2020)
[arXiv:1907.06547 [gr-qc]].


\bibitem{KlinkhamerWang2019-cosm}
F.R.~Klinkhamer and Z.L.~Wang,
\hspace*{0.00mm}``Nonsingular bouncing cosmology from general relativity,''
\textit{Phys. Rev. D} \textbf{100}, 083534 (2019)
[arXiv:1904.09961 [gr-qc]].



\bibitem{KlinkhamerWang2020-pert}
F.R.~Klinkhamer and Z.L.~Wang,
\hspace*{0.00mm}``Nonsingular bouncing cosmology from general relativity:
\hspace*{0.00mm}  Scalar metric perturbations,''
\textit{Phys. Rev. D} \textbf{101}, 064061 (2020)
[arXiv:1911.06173 [gr-qc]].

\bibitem{Wang2020-PhD-thesis}
Z.L. Wang,
\emph{Spacetime Defects and Bouncing Cosmology}, PhD thesis, KIT,
Karlsruhe, Germany, 2020; available from\newline
\verb"https://publikationen.bibliothek.kit.edu/1000126191".


\bibitem{Klinkhamer2014-prd}
F.R.~Klinkhamer,
\hspace*{0.00mm}``Skyrmion spacetime defect,''
\textit{Phys. Rev. D} {\bf 90}, 024007 (2014)
[arXiv:1402.7048 [gr-qc]].


\bibitem{KlinkhamerSorba2014}
F.R.~Klinkhamer and F.~Sorba,
\hspace*{0.00mm}``Comparison of spacetime defects which are homeomorphic
\hspace*{0.00mm}  but not diffeomorphic,''
\textit{J. Math. Phys. (N.Y.)} {\bf 55}, 112503 (2014)
[arXiv:1404.2901 [hep-th]].

\bibitem{Guenther2017}
M. Guenther,
\emph{Skyrmion Spacetime Defect, Degenerate Metric,
and Negative Gravitational Mass},
Master Thesis, KIT, Karlsruhe, Germany, 2017;
available from
\verb"https://www.itp.kit.edu/en/publications/diploma"

\bibitem{Klinkhamer2019-JPCS}
F.R.~Klinkhamer,
\hspace*{0.00mm}``On a soliton-type spacetime defect,''
\textit{J. Phys. Conf. Ser.} {\bf 1275}, 012012 (2019)
[arXiv:1811.01078 [gr-qc]].

\bibitem{Hawking1965}
S.W.~Hawking,
\hspace*{0.00mm}``Occurrence of singularities in open universes,''
\textit{Phys. Rev. Lett.} \textbf{15}, 689 (1965).

\bibitem{Hawking1967}
S.W.~Hawking,
\hspace*{0.00mm}``The occurrence of singularities in cosmology.
\hspace*{0.00mm}   III. Causality and singularities,''
\textit{Proc. Roy. Soc. Lond. A} \textbf{300}, 187 (1967).


\bibitem{HawkingPenrose1970}
S.W.~Hawking and R.~Penrose,
\hspace*{0.00mm}``The singularities of gravitational collapse and cosmology,''
\textit{Proc. Roy. Soc. Lond. A} \textbf{314}, 529 (1970).

\bibitem{Wald1984}
R.M.~Wald,
\emph{General Relativity}
(Chicago University Press, Chicago, USA, 1984).

\bibitem{Horowitz1991}
G.T.~Horowitz,
\hspace*{0.00mm}``Topology change in classical and quantum gravity,''
\textit{Class. Quant. Grav.}  \textbf{8}, 587 (1991).


\bibitem{BoyleFinnTurok2018}
L.~Boyle, K.~Finn, and N.~Turok,
\hspace*{0.00mm}``CPT-symmetric Universe,''
\textit{Phys. Rev. Lett.}  \textbf{121}, 251301 (2018)
[arXiv:1803.08928 [hep-ph]].


\bibitem{Witten1995}
E.~Witten,
\hspace*{0.00mm}``String theory dynamics in various dimensions,''
\textit{Nucl. Phys. B} \textbf{443}, 85 (1995)
[arXiv:hep-th/9503124].


\bibitem{HoravaWitten1996}
P.~Horava and E.~Witten,
\hspace*{0.00mm}``Heterotic and type I string dynamics from eleven dimensions,''
\textit{Nucl. Phys. B} \textbf{460}, 506 (1996)
[arXiv:hep-th/9510209].


\bibitem{Duff1996}
M.J.~Duff,
\hspace*{0.00mm}``M theory (the theory formerly known as strings),''
\textit{Int. J. Mod. Phys. A} \textbf{11}, 5623 (1996)
[arXiv:hep-th/9608117].


\bibitem{IKKT-1997}
N.~Ishibashi, H.~Kawai, Y.~Kitazawa, and A.~Tsuchiya,
\hspace*{0.00mm}``A large-$N$ reduced model as superstring,''
\textit{Nucl. Phys. B} \textbf{498}, 467 (1997)
[arXiv:hep-th/9612115].



\bibitem{Aoki-etal-review-1999} 
H.~Aoki, S.~Iso, H.~Kawai, Y.~Kitazawa, A.~Tsuchiya, and T.~Tada,
\hspace*{0.00mm}``IIB matrix model,''
\textit{Prog. Theor. Phys. Suppl.}  \textbf{134}, 47 (1999)
[arXiv:hep-th/9908038.



\bibitem{KimNishimuraTsuchiya2012}
S.W.~Kim, J.~Nishimura, and A.~Tsuchiya,
\hspace*{0.00mm}``Expanding (3+1)-dimensional universe from a Lorentzian matrix model
\hspace*{0.00mm}  for superstring theory in (9+1)-dimensions,''
\textit{Phys. Rev. Lett.}  \textbf{108}, 011601 (2012)
[arXiv:1108.1540 [hep-th]].



\bibitem{NishimuraTsuchiya2019}
J.~Nishimura and A.~Tsuchiya,
\hspace*{0.00mm}``Complex Langevin analysis of the space-time structure
\hspace*{0.00mm}  in the Lorentzian type IIB matrix model,''
\textit{JHEP} \textbf{1906}, 077 (2019)
[arXiv:1904.05919 [hep-th]].


\bibitem{Witten1979}
E.~Witten,
``The $1/N$ expansion in atomic and particle physics,''
in: G. 't Hooft et. al (eds.),
\emph{Recent Developments in Gauge Theories},
Cargese 1979 (Plenum Press, New York, USA,  1980),
pp. 403--419.



\bibitem{Coleman1985}
S.~Coleman,
``$1/N$,'' in:
\emph{Aspects of Symmetry: Selected Erice Lectures}
(Cambridge University Press, Cambridge, UK, 1985),
Chap. 8.

\bibitem{GreensiteHalpern1983}
J.~Greensite and M.B.~Halpern,
\hspace*{0.00mm}``Quenched master fields,''
\textit{Nucl. Phys. B} \textbf{211}, 343 (1983).


\bibitem{Klinkhamer2020-master}
F.R.~Klinkhamer,
\hspace*{0.00mm}``IIB matrix model: Emergent spacetime from the master field,''
\textit{Prog. Theor. Exp. Phys.} \textbf{2021}, 013B04 (2021)
[arXiv:2007.08485 [hep-th]].


\bibitem{Klinkhamer2020-points}
F.R.~Klinkhamer,
\hspace*{0.00mm}``IIB matrix model: Extracting the spacetime points,''
[arXiv:2008.01058 [hep-th]].


\bibitem{Klinkhamer2020-metric}
F.R.~Klinkhamer,
\hspace*{0.00mm}``IIB matrix model: Extracting the spacetime metric,''
[arXiv:2008.11699 [hep-th]].

\bibitem{Klinkhamer2020-reg-bb-IIB-m-m}
F.R.~Klinkhamer,
\hspace*{0.00mm}``IIB matrix model and regularized big bang,''
Prog. Theor. Exp. Phys. \textbf{2021}, 063B05 (2021)  
[arXiv:2009.06525 [hep-th]].


\bibitem{Klinkhamer2012}
F.R.~Klinkhamer,
\hspace*{0.00mm}``Gravity without curved spacetime: A simple calculation,''
[arXiv:1208.3168 [gr-qc]].


\bibitem{Connes2000}
A.~Connes,
\hspace*{0.00mm}``A short survey of noncommutative geometry,''
\textit{J. Math. Phys.} \textbf{41}, 3832 (2000)
[arXiv:hep-th/0003006].



\bibitem{Steinacker2019}
H.C.~Steinacker,
\hspace*{0.00mm}``On the quantum structure of space-time, gravity,
\hspace*{0.00mm}  and higher spin in matrix models,''
\textit{Class. Quant. Grav.} \textbf{37},  113001 (2020)
[arXiv:1911.03162 [hep-th]].


\bibitem{vanRaamsdonk2020}
M. van Raamsdonk,
\hspace*{0.00mm}``Spacetime from bits,''
\textit{Science} \textbf{370}, 198 (2020).

\bibitem{Das-etal2020a}
S.R.~Das, A.~Kaushal, G.~Mandal, and S.P.~Trivedi,
\hspace*{0.00mm}``Bulk entanglement entropy and matrices,''
\textit{J. Phys. A} \textbf{53},  444002 (2020)
[arXiv:2004.00613 [hep-th]].

\bibitem{Klinkhamer2021-first-look} 
F.R.~Klinkhamer,
``A first look at the bosonic master-field equation of the IIB matrix model,''
[arXiv:2105.05831 [hep-th]].

\bibitem{Klinkhamer2021-solutions-exact}  
F.R.~Klinkhamer,
``Solutions of the exact bosonic master-field equation
from a supersymmetric matrix model,''
[arXiv:2106.07632 [hep-th]].

\end{thebibliography}
\end{document}